\documentclass{ieeeaccess}
\usepackage{cite}
\usepackage{amsmath,amssymb,amsfonts}
\usepackage{algorithmic}
\usepackage{graphicx}
\usepackage{textcomp}

\usepackage{array}
\usepackage{booktabs}
\usepackage{epstopdf}

\newcommand{\tabincell}[2]{\begin{tabular}{@{}#1@{}}#2\end{tabular}}

\begin{document}
\history{Date of publication 2021 00, 0000, date of current version 2021 00, 0000.}
\doi{10.1109/ACCESS.2021.3053427}

\title{Sequential Convolutional Recurrent Neural Networks for Fast Automatic Modulation Classification}

\author{\uppercase{Kaisheng Liao}\authorrefmark{1},
\uppercase{Yaodong Zhao}\authorrefmark{1},
\uppercase{Jie Gu}\authorrefmark{1},
\uppercase{Yaping Zhang}\authorrefmark{1},
\uppercase{Yi Zhong}\authorrefmark{1}
}

\address[1]{Science and Technology on Electronic Information Control Laboratory, Chengdu 610000, China}

\markboth
{Author \headeretal: Preparation of Papers for IEEE TRANSACTIONS and JOURNALS}
{Author \headeretal: Preparation of Papers for IEEE TRANSACTIONS and JOURNALS}

\corresp{Corresponding author: Yi Zhong (e-mail: eric.yi.zhong@outlook.com).}

\begin{abstract}
A novel and efficient end-to-end learning model for automatic modulation classification is proposed for wireless spectrum monitoring applications, which automatically learns from the time domain in-phase and quadrature data without requiring the design of hand-crafted expert features.
With the intuition of convolutional layers with pooling serving as the role of front-end feature distillation and dimensionality reduction, sequential convolutional recurrent neural networks are developed to take complementary advantage of parallel computing capability of convolutional neural networks and temporal sensitivity of recurrent neural networks.
Experimental results demonstrate that the proposed architecture delivers overall superior performance in signal to noise ratio range above -10~dB, and achieves significantly improved classification accuracy from 80\% to 92.1\% at high signal to noise ratio range, while drastically reduces the average training and prediction time by approximately 74\% and 67\%, respectively.
Response patterns learned by the proposed architecture are visualized to better understand the physics of the model.
Furthermore, a comparative study is performed to investigate the impacts of various sequential convolutional recurrent neural network structure settings on classification performance.
A representative sequential convolutional recurrent neural network architecture with the two-layer convolutional neural network and subsequent two-layer long short-term memory neural network is developed to suggest the option for fast automatic modulation classification.
\end{abstract}


\titlepgskip=-15pt

\maketitle

\section{Introduction}
\label{sec:introduction}
\PARstart{W}{ireless} spectrum monitoring over time, space and frequency is important for effective use of the scarce spectral resources in various commercial areas \cite{hoyhtya2016spectrum, barber2004using, yucek2009survey, zheng2018big, thilina2013machine}.
As an integral part of wireless spectrum monitoring systems, automatic modulation classification (AMC) is used to recognize modulation types without prior knowledge of the received signals and channel parameters \cite{weber2015automatic, clancy2007applications, aslam2012automatic}.
AMC has been proven to be an essential capability for transmitter identification, wireless spectrum anomaly detection and radio environment awareness.
It improves radio spectrum utilization and opens the possibility of intelligent decision for context-aware autonomous wireless spectrum monitoring systems.

The existing AMC approaches discussed in literature can be roughly brought down into the following two categories: (i) likelihood-based approaches; and (ii) feature-based approaches \cite{dobre2007survey, nandi1998algorithms}.
For the first category, the likelihood-based approaches utilize hypothesis testing theory and form a judgment criterion by analyzing statistical characteristics of signals \cite{xu2010likelihood, wei2000maximum}.
In likelihood-based approaches, modulation classification is framed as Bayesian estimation to optimize the probability of classification.
However, approaches of this type are not robust in the presence of unknown channel conditions and suffer from heavy computational load on their practical implementations.
Traditional feature-based approaches mainly focus on expert feature extraction and classification criteria \cite{ramkumar2009automatic, wu2008novel, park2008automatic, swami2000hierarchical, hsue1990automatic, soliman1992signal}.
They utilize expert features such as higher order cyclic moments for modulation classification.
It is easy and simple for these approaches to be implemented in practical systems.
However, hand-crafting expert features and hard-coding rules for modulation classification make it difficult to scale to new modulation types in non-cooperative scenarios.

Recently, researchers in wireless communications have started to apply deep neural networks to cognitive radio tasks with some success \cite{o2016convolutional, hong2017automatic, west2017deep, ali2017automatic, lstm2018, o2018over, sadeghi2018adversarial, zhang2018automatic, teng2018polar, sun2018automatic, kulin2018end, tang2018digital, zheng2019fusion, ramjee2019fast, zhang2020automatic}.
The authors in \cite{o2016convolutional, o2018over} demonstrated that convolutional neural networks (CNNs) trained on time domain in-phase and quadrature (IQ) data significantly outperform conventional expert feature-based approaches.
The authors in \cite{lstm2018, hong2017automatic} utilized recurrent neural networks (RNNs) for learning temporal representations to achieve higher classification accuracy than that of the CNNs introduced in \cite{o2016convolutional}.
In \cite{west2017deep}, the authors directly adopted convolutional long short-term deep neural networks (CLDNNs) from voice processing domain.
The authors in \cite{wang2019data} developed a data-driven fusion method to obtain better classification accuracy using the combination of the two CNNs trained on different datasets.
Ramjee et al. \cite{ramjee2019fast} performed a comparative study of various typical deep neural networks and reduced the training complexity by reducing the input dimensionality with subsampling techniques.

In autonomous wireless spectrum monitoring systems, online learning is fundamental for accommodating new emerging modulation types and complex environmental circumstances.
Nevertheless, those RNN models delivering high classification accuracy suffer from computational complexity and long training time.
In this work, we develop a novel and efficient sequential convolutional recurrent neural network (SCRNN) architecture combining parallel computing capability of CNNs with temporal sensitivity of RNNs.
Experimental results demonstrate that our approach outperforms the state-of-the-art on classification performance, while significantly improves the rate of convergence compared with the CNN and RNN alone architectures.

The rest of the paper is organized as follows. In Section \ref{sec: dataset_baseline}, an overview of the modulation benchmark dataset is introduced, and the two baseline models are briefly explained. The proposed model and the parameters used for training along with other implementation details are clearly stated in Section \ref{sec: proposed_model}. Section \ref{sec: results_discussion} details the classification results and discusses the advantages of the proposed model. Conclusions and future work are presented in Section \ref{sec: conclusion}.

\section{Dataset and Baselines}
\label{sec: dataset_baseline}

\subsection{Dataset}
In a wireless spectrum monitoring system, the received signal can be typically represented as:

\begin{equation}\label{equ:signal}
  r(t)=s(t)*h(t)+n(t)
\end{equation}

where $s(t)$ denotes the noise free complex baseband envelope of the received signal, and $h(t)$ refers to the time varying impulse response of the transmitted wireless channel. $n(t)$ represents the additive white Gaussian noise (AWGN) reflecting thermal noise. The complex received signal $r(t)$ is commonly sampled in IQ format due to its simplicity.
% for radio hardware design

A typical modulation dataset RadioML2016.10a generated by GNU Radio is used as the benchmark dataset for training and evaluating the performance of the proposed architecture, similar as the MNIST dataset in the vision domain \cite{o2016radio}.
The dataset follows the signal representation as given in equation \ref{equ:signal}.
Detailed parameter description of the dataset is shown in Table \ref{tab:dataset}.
Radio channel effects are relatively well characterized in the dataset.
Chanel imperfections such as multi-path fading, random walk drifting of carrier frequency oscillator and sample time clocks, AWGN, along with unknown scale, translation, and dilation transformation are introduced into the signal in the dataset for reflecting the real electromagnetic environment \cite{o2016radio}.
The dataset is labeled with both signal to noise ratio range (SNR) ground truth and modulation types.

\begin{table}[!t]
\renewcommand{\arraystretch}{1.3}
\caption{Benchmark dataset parameters.}
\label{tab:dataset}
\centering
\begin{tabular}{|l|l|}
\hline
Dataset & RadioML2016.10a\\
\hline
Modulations & \tabincell{l}{ 8 Digital Modulations: BPSK,\\ QPSK, 8PSK, 16QAM, 64QAM,\\ BFSK, CPFSK, and PAM4\\ 3 Analog Modulations: WBFM,\\ AM-SSB, and AM-DSB}\\
\hline
Length per sample & 128 \\
\hline
Signal format & In-phase and quadrature (IQ) \\
\hline
Signal dimension & 2$\times$128 per Sample \\
\hline
Duration per sample & 128 $\mu$s \\
\hline
Sampling frequency & 1~MHz \\
\hline
Samples per symbol & 8 \\
\hline
SNR Range & [-20~dB, -18~dB, -16~dB, $\ldots$, 18~dB] \\
\hline
Total number of samples & 220000 vectors\\
\hline
Number of training samples & 198000 vectors \\
\hline
Number of test samples & 22000 vectors \\
\hline
\end{tabular}
\end{table}

\subsection{Baselines}

The two models are chosen as the baselines for further comparisons due to their results showing the significant improvements upon expert feature-based approaches. Any further improvements should be considered state-of-the-art.

\begin{figure}[!t]
\centering
\includegraphics[width=0.48\textwidth]{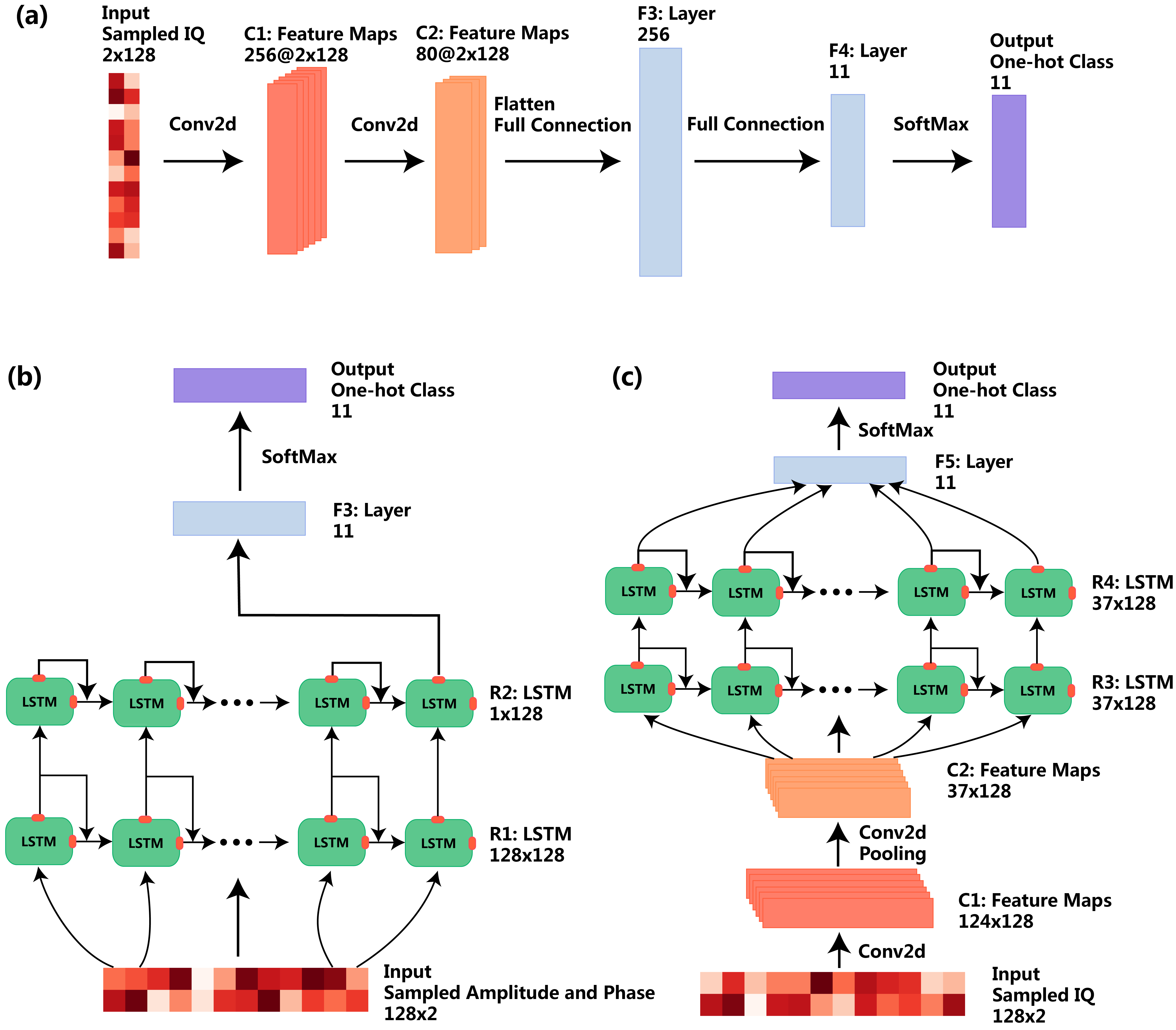}
\caption{Schematic diagram of the (a) convolutional neural network (CNN) baseline model, (b) long short-term memory (LSTM) baseline model, and (c) proposed sequential convolutional recurrent neural network (SCRNN) model.}
\label{fig:model}
\end{figure}

One is the CNN architecture proposed by O'shea et al. \cite{o2016convolutional}.
As shown in Fig.~\ref{fig:model}(a), the baseline model is a 4-layer network made up of two convolutional layers and two dense layers.
Each hidden layer utilizes rectified linear unit (ReLU) activation functions and dropout of 50\% except for a softmax activation function on the one-hot output layer.
Adam optimizer and categorical cross entropy loss function are applied to the base model.

The other baseline model is proposed by Rajendran et al. \cite{lstm2018}, shown in Fig.~\ref{fig:model}(b).
The model is comprised of two 128-unit long short-term memory (LSTM) layers and an 11-unit dense layer with a softmax activation.
The first LSTM layer returns the full sequences while the second one just returns the last state.
The dropout is also adopted to reduce overfitting.
Adam optimizer and categorical cross entropy loss function are applied to the model.
Note that this model learns from the time domain information of the modulation schemes using amplitude-phase format, instead of IQ format.

\section{Sequential Convolutional Recurrent Neural Networks}
\label{sec: proposed_model}

\subsection{Motivation}

Generally, the received radio signals sampled at discrete time steps are of time domain sequences.
In \cite{lstm2018}, a two-layer LSTM architecture is proposed and achieves a good classification accuracy of 86\% at high SNRs.
However, these models using RNNs suffer from much slower training time than that of the CNNs, due to their computational complexity and unparallel computing capability.
Thus, a new novel and efficient SCRNN architecture is proposed with the combination of the speed and lightness of CNNs and the temporal sensitivity of RNNs. 
Furthermore, as a variant of RNN, LSTM is adopted instead of simple RNN in the proposed architecture to remember long-term dependencies and avoid the gradient vanishing problem.
In SCRNN architectures, the convolutional layers with pooling acting as the role of front-end feature distillation and dimensionality reduction turn the long input sequences into much shorter representations of high-level features, which then become the input for subsequent LSTM layers to learn long-term temporal coherence of modulations.

\subsection{Model Description}

Fig.~\ref{fig:model}(c) provides the illustration of the proposed SCRNN architecture.
As schematically shown in Fig.~\ref{fig:model}(c), the first and second convolutional layers each contain 128 5-tap filters except for the first one followed by a max-pooling layer with a pooling size of 3.
The layer 3 and layer 4 are LSTM layers composed of 128 units each, and both return the full sequences.
The last dense layer contains 11-class neurons representing the modulation schemes.

ReLU activation functions are applied to the convolutional and LSTM layers.
The last dense layer utilizes a softmax activation to achieve modulation classification.
Dropout regularization combined with max norm has been proven to be of better performance for preventing overfitting.
Categorical cross entropy is adopted as the loss function, which can be written as:
\begin{equation}\label{equ:crossentropy}
  L = -\frac{1}{N}\sum_{i=1}^{N}\mathbf{y_{i}} \cdot log(\mathbf{\hat{y}_{i}})
\end{equation}
where $\mathbf{y_{i}}$ represents the ground truth in the form of one-hot encoding, and $\mathbf{\hat{y}_{i}}$ refers to the prediction. $N$ denotes the training batch size. 
Adam optimizer with a learning rate of 0.001 is utilized due to its computational efficiency.

\subsection{Implementation details}
The total 220000 samples in the RadioML2016.10a dataset are split into two, one training set of 198000 (90\%) samples and the other test set of 22000 (10\%) samples. 
The dataset is split equally among all considered modulation types using the stratified sampling strategy. 
Instead of extracting the amplitude and phase features of the signals manually in advance \cite{lstm2018}, we adopted IQ components as input directly. 
A batch size of 128 is used on each training epoch and the early stop strategy is adopted.

All training and prediction are implemented in Keras libarary \cite{chollet2018keras} on the backend of TensorFlow \cite{abadi2016tensorflow}. The Nvidia Cuda enabled Tesla K80 is used to speed up the calculation.

\section{Results and Discussion}
\label{sec: results_discussion}

The classification performance of the models on the benchmark dataset is discussed in this section.
We inspect and compare the classification accuracy and rate of convergence between the baseline models and the proposed SCRNN model.
In addition, the varying kernel sizes, kernel types and layer depths are further investigated to find the optimal SCRNN architecture.

\begin{figure}[htbp]
\centering
\includegraphics[width=0.42\textwidth]{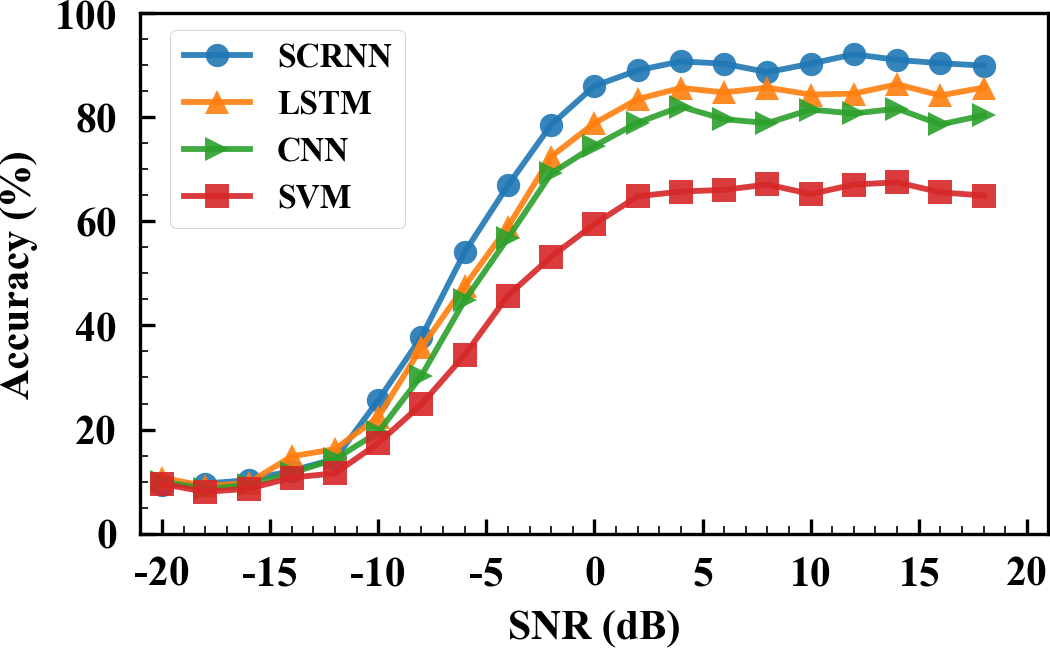}
\caption{Classification accuracy comparison of the proposed SCRNN model with others on the benchmark dataset.}
\label{fig:accuracy_comparison}
\end{figure}

The classification accuracy of all the models are presented in Fig. \ref{fig:accuracy_comparison}.
It can be seen that the proposed SCRNN model delivers a significantly improved accuracy of 92.1\% at high SNRs.
The CNN and LSTM model as baselines are compared to the proposed SCRNN model.
It shows that the SCRNN model consistently achieves higher accuracy than the other two baselines in the SNR range from $-10$~dB to 18~dB, and significantly outperforms the CNN baseline model by 12\% and the LSTM baseline model by 6\% improvement at high SNRs.
Additionally, it is observed that the proposed SCRNN model achieves exceeding performance than that of the CNN and LSTM baseline models in the SNR range from $-10$~dB to 0~dB, where the two baseline models behave nearly the same.
It implies that the convolutional layers of the SCRNNs serving as the role of feature distillation boost the learning ability of the temporal features under low SNR circumstances.
The traditional support vector machine (SVM) approach showing relatively poor classification performance is also summarized in Fig. \ref{fig:accuracy_comparison} for comparison.
Note that all models are fed with the same training and test data of IQ format for this comparison except for the LSTM model with amplitude-phase format, and the standardization instead of $L_2$ normalization in \cite{o2016convolutional} is adopted to scale the input for all models.

\begin{figure}[htbp]
\centering
\includegraphics[width=0.48\textwidth]{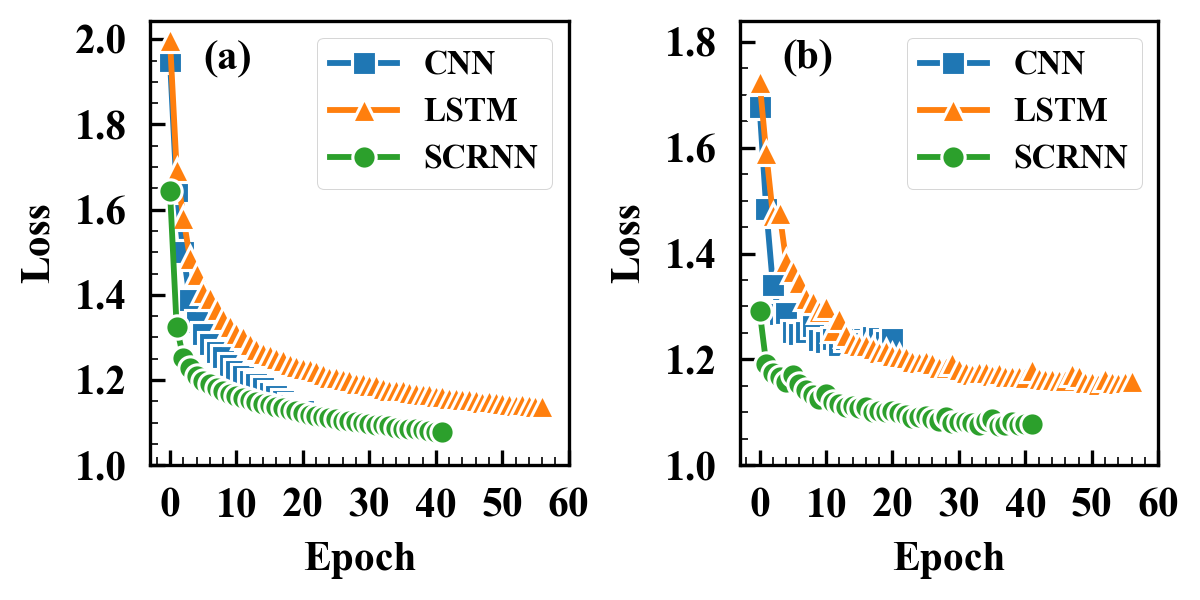}
\caption{Training history including the (a) training loss and (b) validation loss between the baseline models and the proposed SCRNN model.}
\label{fig:training_history_comparison}
\end{figure}

\begin{table}[bt]
\renewcommand{\arraystretch}{1.3}
\caption{Training and prediction time comparison between the two baseline models and the SCRNN model.}
\label{tab: training_time_comparison_with_baselines}
\centering
\begin{tabular}{llll}
\hline
Models & CNN\cite{o2016convolutional} & LSTM\cite{lstm2018} & SCRNN\\
\hline
\tabincell{l}{Number of trainable parameters} & 5369947 & 200075 & 398731\\
\tabincell{l}{Training time per epoch (s)} & 30 & 800 & 280\\
\tabincell{l}{Training epochs} & 20 & 56 & 41\\
\tabincell{l}{Training time (s)} & 600 & 44800 & 11480\\
\tabincell{l}{Prediction time ($\mu$s/sample)} & 1000 & 2000 & 661\\
\hline
\end{tabular}
\end{table}

Fig. \ref{fig:training_history_comparison} shows the training history including the (a) training loss and (b) validation loss compared between the baseline models and the proposed SCRNN model.
According to the training history, the LSTM baseline model achieves the second less loss value but remains the lowest rate of convergence; the CNN baseline model obtains faster rate of convergence but yields the largest loss value, while the proposed SCRNN model retains the fastest rate of convergence and achieves the least loss value among the three.
The average training and prediction time together with the network size of the three models are compared in Table \ref{tab: training_time_comparison_with_baselines}.
It can be seen that though the introduction of the convolutional layers in the SCRNN leads to nearly double the network size, the average training and prediction time of the proposed SCRNN model are drastically reduced to only 280 seconds per epoch and 661~$\mu$s per sample respectively, compared to 800 seconds per epoch and 2000~$\mu$s per sample of the LSTM model.
These are fairly consistent with the insight that the convolutional layers with pooling before RNN serve as the role of feature distillation and dimensionality reduction, analogous to front-end matched filters, synchronizer and sampler for temporal features in typical wireless systems.
Thus, the improved quality of the input for the SCRNN model makes it significantly reduce the training time and achieve the fastest prediction time.

\begin{figure}[htbp]
\centering
\includegraphics[width=0.48\textwidth]{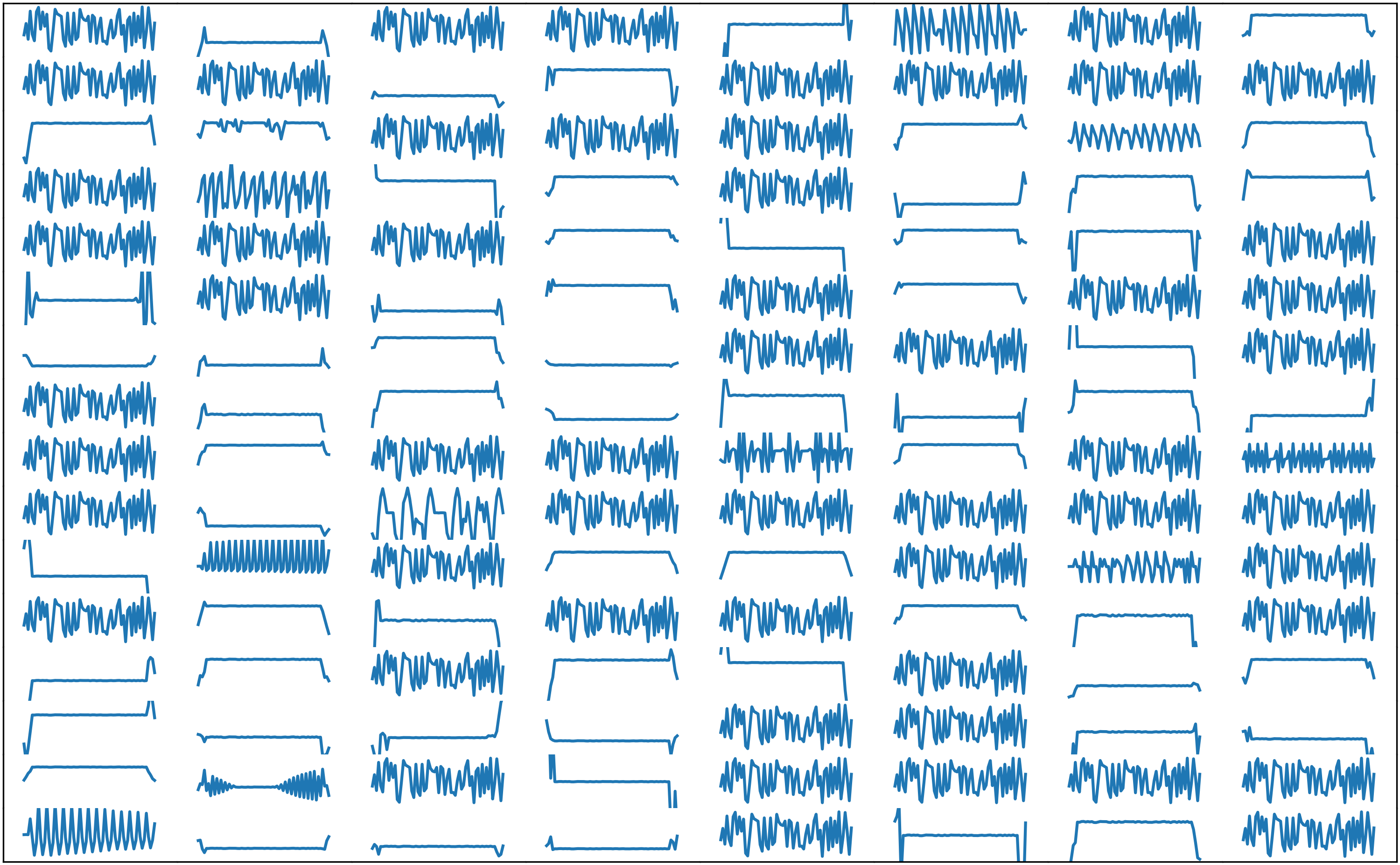}
\caption{Response patterns of the 128 filters learned by the first convolutional layer of the SCRNN.}
\label{fig:conv_viz}
\end{figure}

To gain intuition on what convolution layers are learning in SCRNN architectures, the response patterns of the 128 filters learned by the first convolutional layer are illustrated in Fig. \ref{fig:conv_viz}, showing that some filters encode expert-like patterns (i.e. BPSK-like pattern in row 1 column 6) and others even encode more complicated patterns.
It further confirms that the convolutional layers of the SCRNNs act as the role of front-end feature distillation with coherent features refined and redundant features filtered out, enabling the improved rate of convergence.

\begin{figure}[htbp]
\centering
\includegraphics[width=0.48\textwidth]{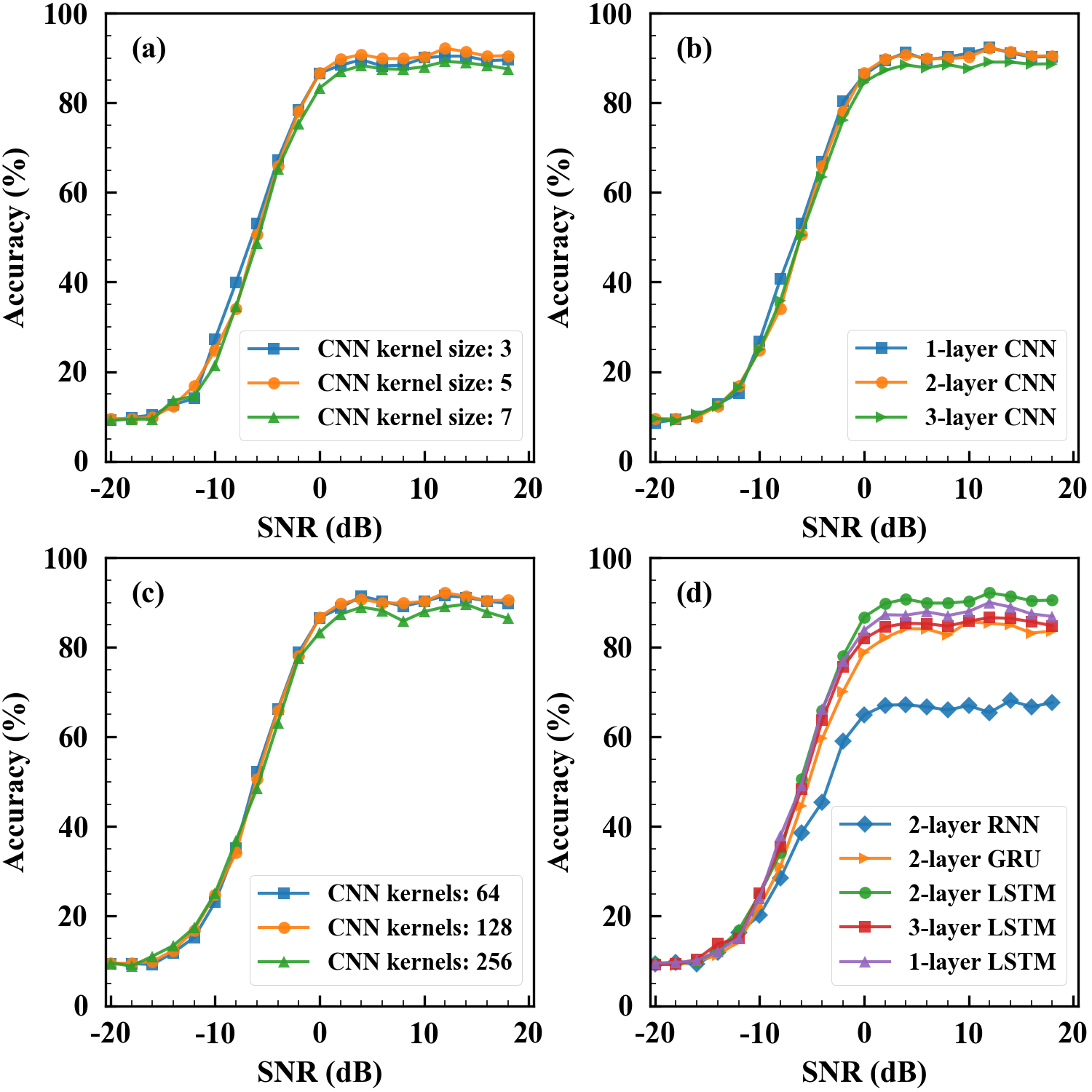}
\caption{Classification performance of different SCRNN structure configurations for varying (a) CNN kernel sizes, (b) CNN layer depths, (c) CNN kernel numbers, (d) RNN types and layer depths.}
\label{fig:different_architecture_SCRNN}
\end{figure}

To gain more insight into the SCRNN architecture, we further investigate the effects of various SCRNN structure settings varying CNN kernel sizes, CNN layer depths, CNN kernel numbers, RNN types and RNN layer depths on classification performance.

As shown in Fig. \ref{fig:different_architecture_SCRNN}(a), varying the CNN kernel sizes of the SCRNN has minimal impact on classification performance.
The architecture with kernel size of 5 produces slightly better classification accuracy than others in SNR range from 0~dB to 18~dB, while the architecture with kernel size of 3 leads to marginally higher classification accuracy in SNR range from -10~dB to -6~dB.
The kernel size of 5 is used for the remaining experiments.

Increasing of the CNN layer depths with pooling reduces the input dimensionality for subsequent LSTM layers in the SCRNN architecture, and hence reduces the training time.
Fig. \ref{fig:different_architecture_SCRNN}(b) proves that the input dimensionality reduction shows very limited effects on classification performance.
However, the performance of the LSTM baseline model starts to decay significantly when reducing the input dimensionality \cite{ramjee2019fast}.
It is implied that the SCRNN architecture is much more robust to dimensionality reduction. 
Thus, it makes possible for deploying online learning model on autonomous wireless spectrum monitoring systems.

Fig. \ref{fig:different_architecture_SCRNN}(c) provides that the 64-kernel and 128-kernel structures deliver the very similar performance, while the performance of 256-kernel structure starts to drop due to the overfitting.
Fig. \ref{fig:different_architecture_SCRNN}(d) shows the different settings of the RNN types and layer depths in the SCRNN architecture.
It can be observed that the performance of the LSTM type is apparently superior to that of the gated recurrent unit (GRU) and the simple RNN type.
Experimental results of varying LSTM layer depths suggest that the 2-layer LSTM of the SCRNN achieves the best classification accuracy.
Therefore, the optimal SCRNN architecture with the 2-layer CNN and subsequent 2-layer LSTM is recommended for online learning.

\begin{figure}[htbp]
\centering
\includegraphics[width=0.48\textwidth]{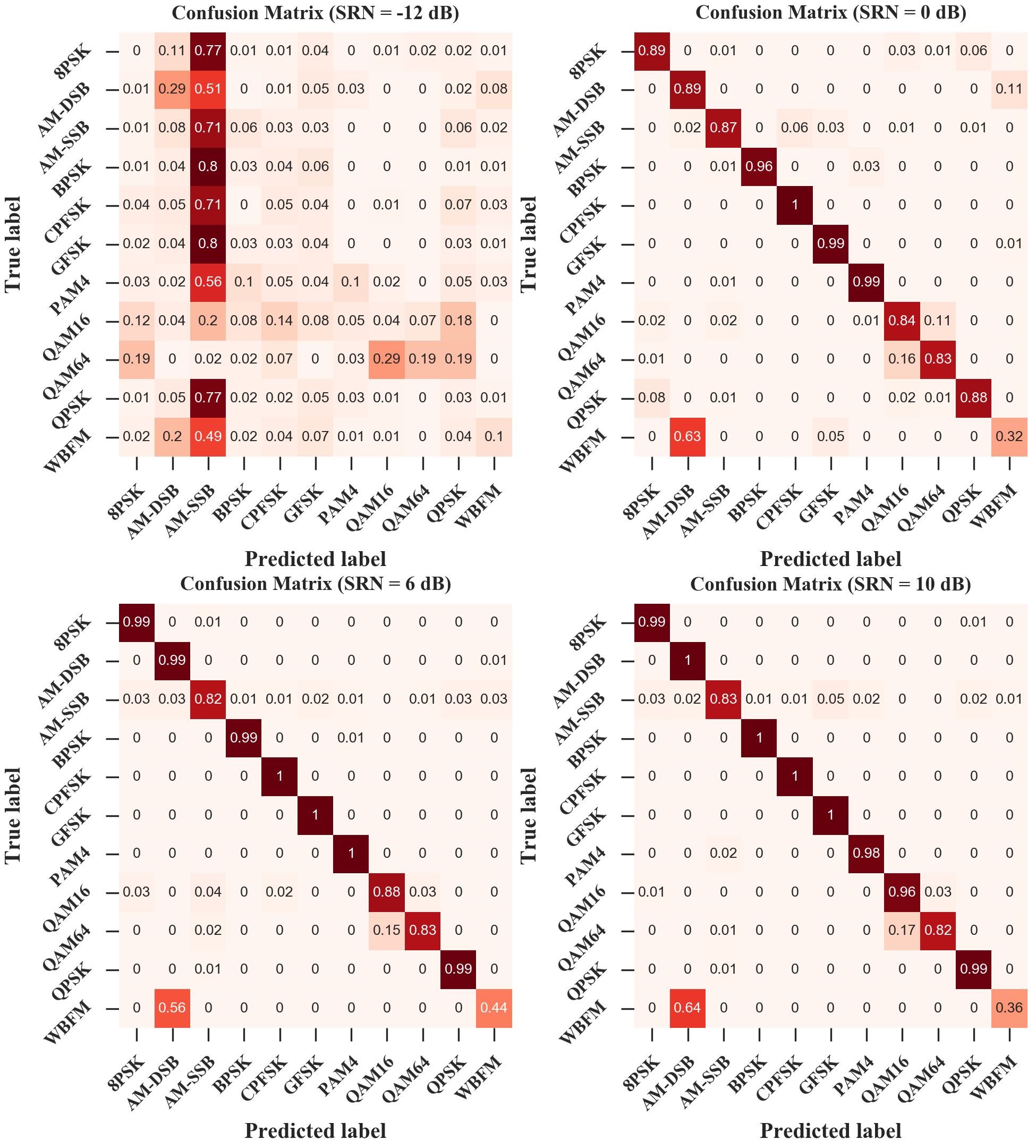}
\caption{Confusion matrices for the optimized SCRNN architecture on the benchmark dataset at various SNRs.}
\label{fig:confusion_matrix}
\end{figure}

To evaluate how classification performance varies with SNRs, confusion matrices of the optimal SCRNN model at various SNRs are investigated. 
For a confusion matrix, each column represents the predicted modulation type and each row represents the real modulation type. 
The numerical value on each grid denotes the prediction probability of the corresponding modulation type.

As illustrated in Fig. \ref{fig:confusion_matrix}, the diagonals become gradually sharper with increasing SNR, yet two primary confusions exist even at high SNRs. 
One is among the analog modulations. 
This is mainly due to the silent period existing in the analog audio signal \cite{o2016convolutional}. 
The other is between QAM16 and QAM64 as the former is a subset of the latter.

\section{Conclusion}
\label{sec: conclusion}

In this paper, a novel and efficient SCRNN architecture for AMC has been developed.
Compared with the CNN and LSTM baseline models, the proposed architecture takes full advantage of the complementarity of CNNs and RNNs.
Thus, it makes the classification accuracy deliver the state-of-the-art performance, improved from 80\% to 92.1\% at high SNRs.
The average training and prediction time of the proposed architecture are significantly reduced by approximately 74\% and 67\% respectively, paving the way for deployment of online learning models on autonomous wireless spectrum monitoring systems.
Response patterns learned by the proposed architecture have been investigated to better understand what feature pattern each filter in the convolutional layers is receptive to.
Additionally, a comparative study of various structure settings of SCRNNs has been performed, and a representative SCRNN architecture with the 2-layer CNN and subsequent 2-layer LSTM was developed to recommend for fast AMC.
Future work will focus on validation on radio signals with varying symbol rates and bandwidths.
Second, unsupervised or deep reinforcement learning approaches for AMC should be investigated due to the lack of necessary signal labels in real wireless spectrum monitoring systems.
Finally, stream learning without requiring to retrain the entire network from scratch is also a worthy direction for future research.

\ifCLASSOPTIONcaptionsoff
  \newpage
\fi

\bibliographystyle{plain}
\bibliography{scrnn_paper}

\EOD

\end{document}